\documentclass[11pt]{article}
\usepackage{moriond,epsfig}

\def\Journal#1#2#3#4{{#1} {\bf #2}, #3 (#4)}

\def\NIMA{{\em Nucl. Instrum. Methods} A}

\def\PLB{{\em Phys. Lett.}  B}
\def\PRL{\em Phys. Rev. Lett.}
\def\PRD{{\em Phys. Rev.} D}

\def\EPJC{{\em Eur. Phys. Jour.} C}

\def\be{\begin{equation}}
\def\ee{\end{equation}}
\def\bea{\begin{eqnarray}}
\def\eea{\end{eqnarray}}

\newcommand{\phistar}{\mbox{$\phi^{*}_{\eta}$}}

\begin{document}

FERMILAB-CONF-11-227-E-PPD
MAN/HEP/2011/07

\vspace*{4cm}

\title{Single W and Z boson production properties and asymmetries}

\author{ Mika Vesterinen }

\address{The School of Physics and Astronomy, The University of Manchester, \\
Oxford Road, Manchester, M13 9PL, England.}

\maketitle\abstracts{
Recent analyses of single $W$ and $Z$ boson production properties and asymmetries
from the CDF and D\O\ experiments at the Fermilab Tevatron are reported.
For $W$ boson production, measurements of the production and lepton charge
asymmetries are presented.
For $Z/\gamma^*$ production, the following measurements are presented:
$d\sigma/dy$, $(1/\sigma) (d\sigma/dp_T)$, $(1/\sigma) (d\sigma/d\phistar)$,
lepton angular coefficients, and $A_{FB}$ with extraction of $\sin^2\theta_W$
and the light quark couplings to the $Z$.
Most of these measurements are in good agreement with QCD predictions.
}

\section{Introduction}

Production of electroweak vector bosons at hadron colliders
provides a rich testing ground for predictions of the Standard Model.
The production cross sections and distributions are sensitive 
to higher order QCD corrections, 
and to the parton distribution functions (PDFs).
Leptonic (involving electrons and muons rather than taus) final states are experimentally convenient,
due to the relatively low background rates
and straightforward triggering on single (or pairs of) high transverse momentum, $p_T$,
leptons.

\section{\boldmath{$W$} boson charge asymmetry}

\begin{figure}
\centering
\includegraphics[width=0.4\textwidth]{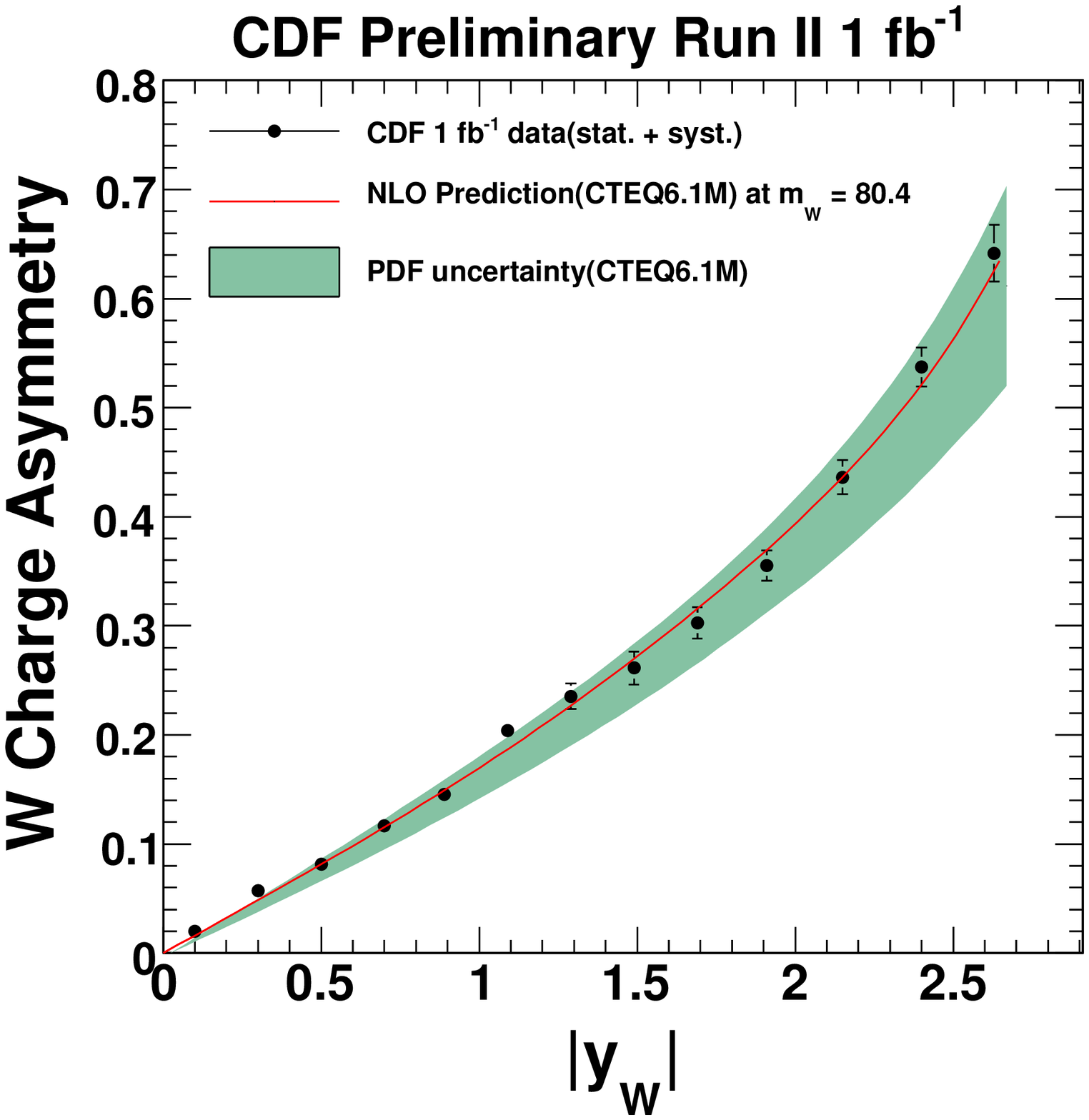}
\includegraphics[width=0.4\textwidth]{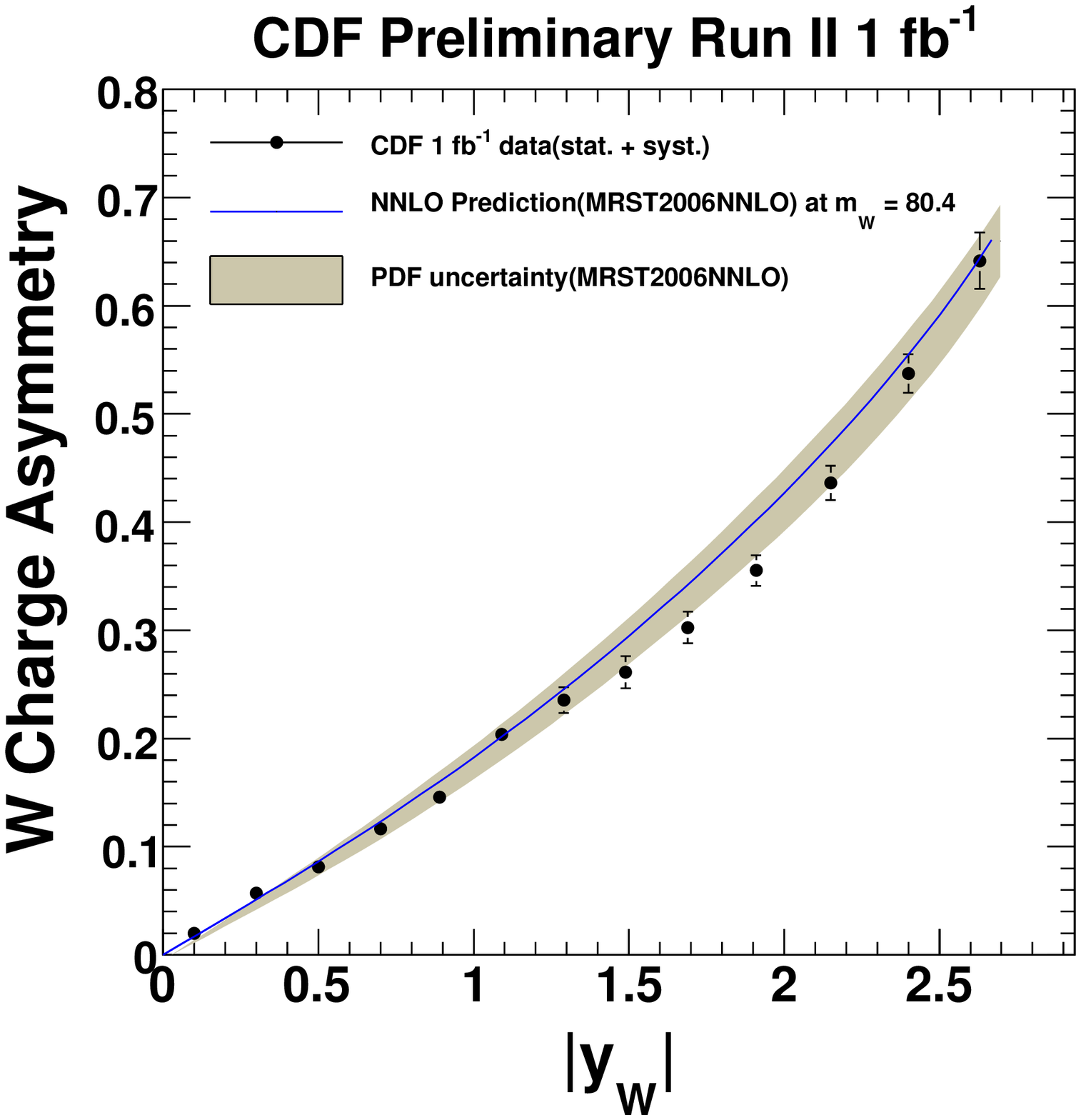}
\label{Fig:CDF_W_charge_asy}
\caption{Comparison of the measured $W$ boson charge asymmetry from CDF
with (left) a NLO QCD prediction with CTEQ 6.6 PDFs, and (right)
a NNLO QCD prediction with MRST2008 PDFs.}
\end{figure}

The production of $W$ bosons at the Tevatron is mostly via the annihilation
of valence light quarks; for example the annihilation
of a $u$ from a proton with a $\bar{d}$ from an antiproton to produce a $W^+$.
It is well known that $u(\bar{u})$ quarks tend to carry a larger fraction ($x$) of the $p(\bar{p})$
momentum than $d(\bar{d})$ quarks.
For the process $p\bar{p} \rightarrow W$, this implies a preferred boost of $W^+$s along
the {\em proton} direction, and along the {\em antiproton} direction for $W^-$s.
The $W$ boson production asymmetry is defined as
\[ A(y_W) = \frac{ N^+(y_W) - N^-(y_W)}{N^+(y_W) + N^-(y_W) } \]
where $y_W$ is the boson rapidity,
and is primarily sensitive to the slope of the ratio of $u$ and $d$ quark PDFs
as a function of $x$.
Unfortunately, $y_W$ is unobservable due to the unknown momentum of the
neutrino along the beam direction.
A novel solution suggested by Bodek {\it et al}~\cite{A_W_method} involves
constraining the invariant mass of the charged lepton and neutrino to the known
$W$ boson mass, leaving two solutions for $y_W$.
Each of these is assigned a weight assuming the known $V-A$ structure of the weak decay vertex.
This method was employed in a measurement by CDF in the $W\rightarrow e \nu_e$ channel
using 1 fb$^{-1}$ of data~\cite{CDF_A_W}.
Figure~\ref{Fig:CDF_W_charge_asy} shows that the measured $A(y_W)$ agrees well
with QCD predictions at both NLO and NNLO accuracies.

\begin{figure}
\centering
\includegraphics[width=0.4\textwidth]{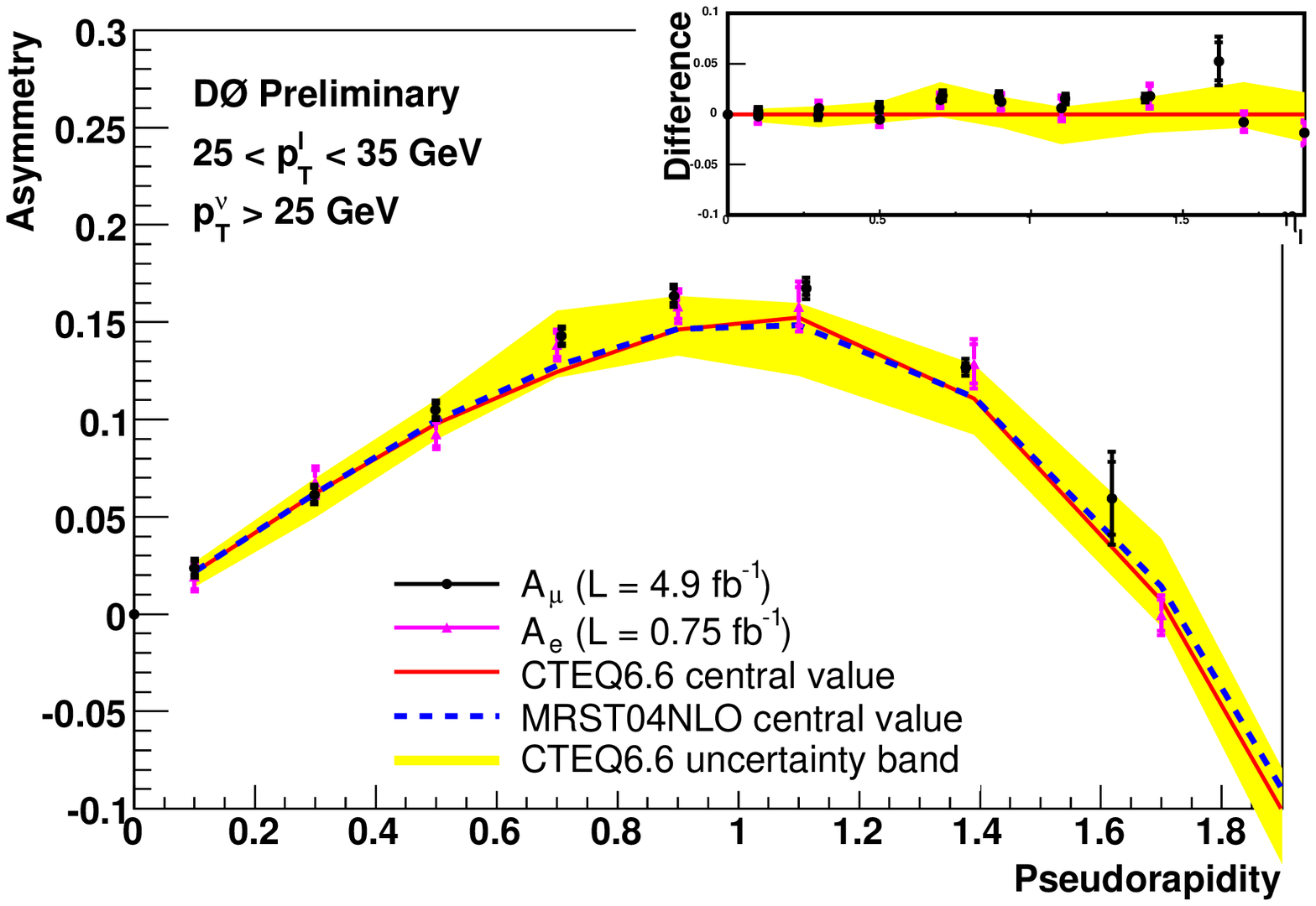}
\includegraphics[width=0.4\textwidth]{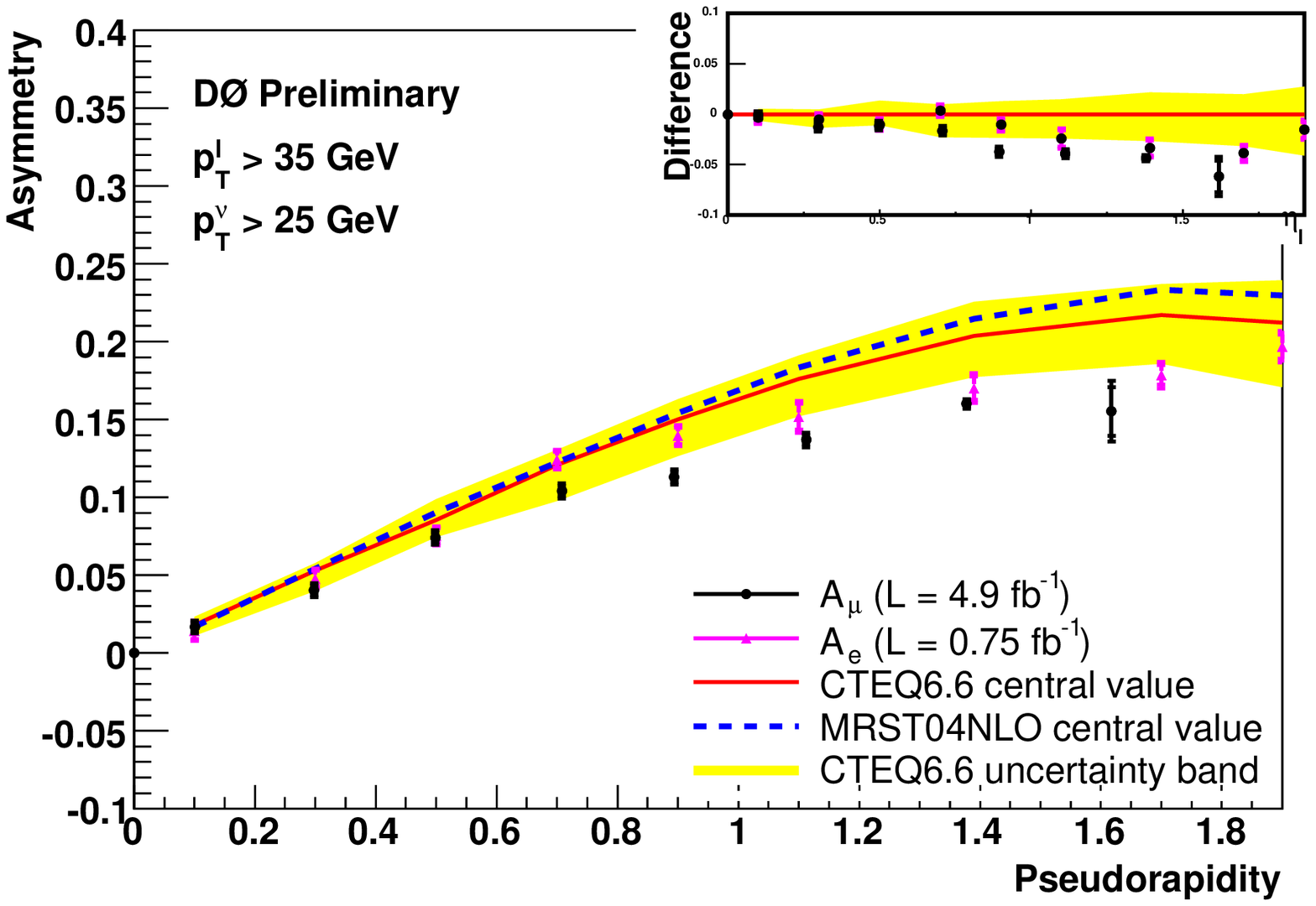}
\label{Fig:D0_muon_charge_asy}
\caption{Comparison of the measured muon charge asymmetry from D\O\
with NLO QCD predictions for (left) $20 > p_T > 35$ GeV,
and (right) $p_T > 35$ GeV.}
\end{figure}

An alternative approach is to measure the asymmetry as a function of the 
{\em observable} lepton pseudorapidity, $\eta_l$.
Unfortunately, the lepton charge asymmetry, $A(\eta_l)$,
is less sensitive to the production asymmetry and thus also the PDFs.
The $V-A$ structure of the decay vertex implies that the charged lepton
tends to head backwards in the $W$ boson rest frame,
i.e. cancelling the production asymmetry;
particularly at low lepton $p_T$ and/or large lepton $\eta$.
Nevertheless, the two approaches provide complementary information.
The D\O\ Collaboration recently measured $A(\eta_l)$ using 4.9 fb$^{-1}$ 
of data in the $W \rightarrow \mu \nu_{\mu}$ channel~\cite{Dzero_A_mu},
and compared to NLO QCD predictions as shown in figure~\ref{Fig:D0_muon_charge_asy}.
The measurement is performed in two bins of muon $p_T$ which
partially disentangles the production and decay asymmetries.
Interestingly, this measurement does not agree so well with the
QCD predictions, particularly at larger muon $p_T$ and pseudorapidity.

\section{\boldmath{$Z/\gamma^*$} rapidity distribution}

\begin{figure}
\centering
\includegraphics[width=0.45\textwidth]{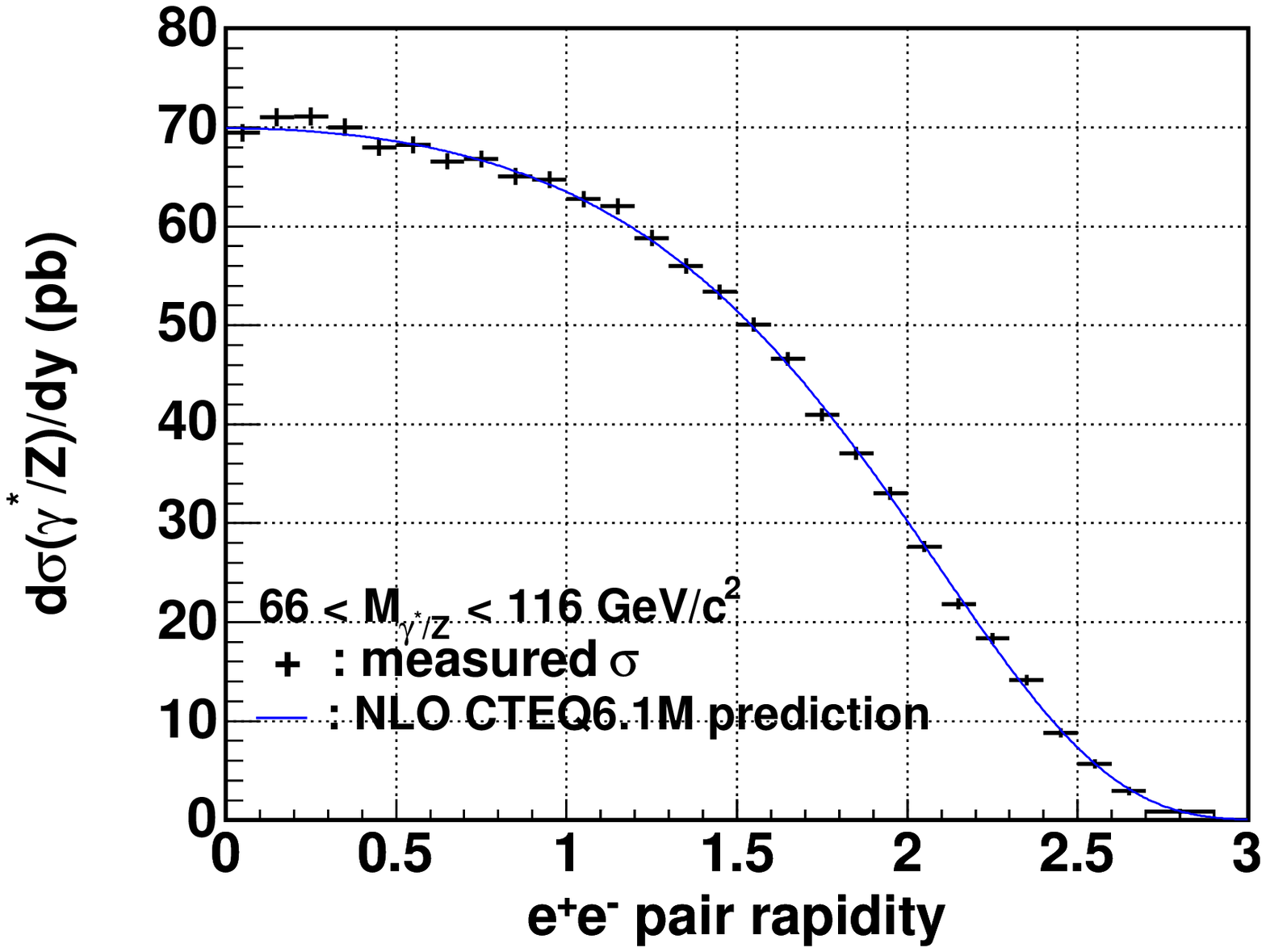}
\includegraphics[width=0.45\textwidth]{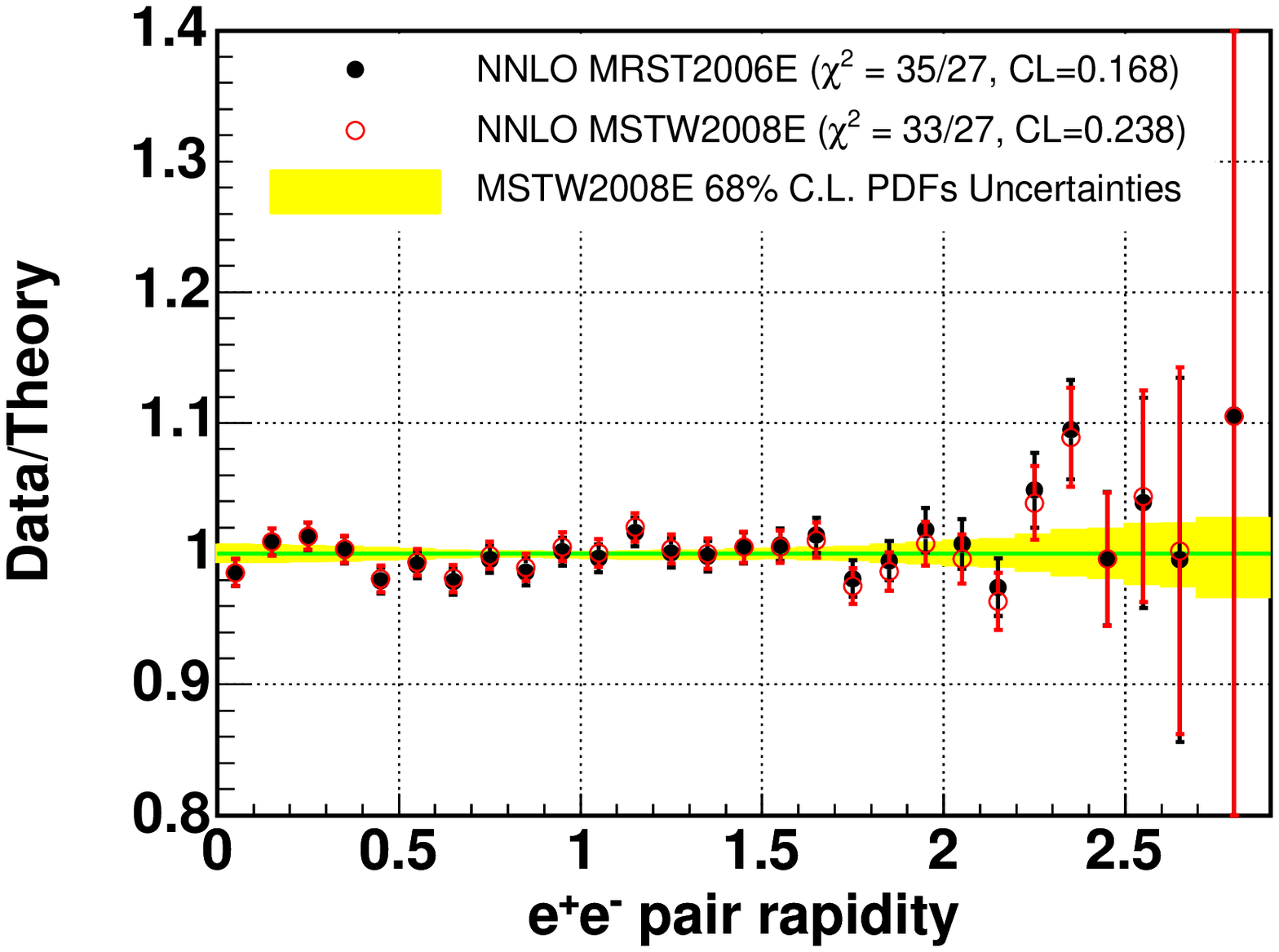}
\label{Fig:CDF_rapidity}
\caption{Left: measured $d\sigma/dy$ from CDF compared to a NLO QCD prediction.
Right: ratio of the data to NNLO QCD predictions, where the yellow band represents
the uncertainty on the prediction due to the PDFs.
}
\end{figure}

The rapidity, $y$, of the dilepton system in $Z/\gamma^*$ decays
is directly related to the $x$ of the two partons:
$x_{1,2} = (M_{ll}/\sqrt{s})e^{\pm y}$, where
$M_{ll}$ is the dilepton invariant mass, and $\sqrt{s}$ is the centre
of mass energy of the collider.
Events with large rapidity correspond to the annihilation of a low-$x$ parton
and a high-$x$ parton.
Thus, a measurement of $d\sigma/dy$ provides additional information on the PDFs
that is complementary to the $W$ charge asymmetry.
CDF has measured $d\sigma/dy$ in the $e^+e^-$ decay channel using 1 fb$^{-1}$ of data~\cite{CDF_dsigma_dy}.
Figure~\ref{Fig:CDF_rapidity} shows that the data are in good agreement with NLO/NNLO QCD predictions,
over the full range of probed rapidities.
The $Z/\gamma^*$ production cross section is measured as 257 $\pm$ 16 pb, also in agreement with NLO/NNLO QCD predictions.

\section{\boldmath{$Z/\gamma^*$} transverse momentum distribution}

At lowest order in $Z/\gamma^*$ production, the dilepton system has zero momentum transverse to the beam direction, $p_T$.
Higher order QCD corrections include radiation of gluons from the one or both of the annihilating quarks.
Alternatively, one or both of the annihilating quarks can result from a gluon splitting into
a pair of quarks.
In addition, the partons may carry some intrinsic transverse momentum within the colliding hadrons.
A good understanding of these effects is paramount for many physics analyses at hadron colliders;
for example the $W$ boson mass measurement, which relies on a precise prediction of the lepton
kinematics for different mass hypotheses.

The D\O\ Collaboration has recently measured the shape of the $p_T$ distribution in the $\mu^+\mu^-$ final
state using 1.0 fb$^{-1}$ of data~\cite{Dzero_mumu_pt}.
For $p_T$ $>$ 10 GeV, NLO QCD is able to describe the data reasonably well,
whilst resummation is needed at lower $p_T$, as implemented at approximate leading-log (LL) in various Monte Carlo event generators,
and at next-to-LL in the ResBos program~\cite{ResBos}.
Compared to the data, ResBos underestimates the cross section for larger $p_T$ ($p_T$ $>$ 50 GeV),
and varying levels of agreement are observed for the different event generators.



This and other recent measurements of the $Z/\gamma^*$ $p_T$ distribution have been dominated
by uncertainties in correcting for detector resolution and efficiency.
An alternative approach is to measure the distribution of a variable that is less sensitive to these
effects, such as $a_T$~\cite{aT_NIM}, or more recently $\phi^*$~\cite{phistar_EPJ} defined as
$\phistar = \tan([(\pi-\Delta\phi)/2]\sin\theta^*$,
where $\Delta\phi$ is the azimuthal opening angle between the two leptons,
and $\cos\theta^* = \tanh[(\eta^{(-)} - \eta^{(+)})/2]$,
with $\eta^{(-)}$ being the pseudorapidity of the negatively charged lepton.
The variable \phistar\ is sensitive to the same physics as the $p_T$,
but is determined exclusively from lepton angles resulting in far better
experimental resolution.
Furthermore, \phistar\ is less correlated than the $p_T$, with efficiencies
of typical $Z/\gamma^*$ event selection requirements; e.g. on lepton isolation.

\begin{figure}
\centering
\includegraphics[width=0.7\textwidth]{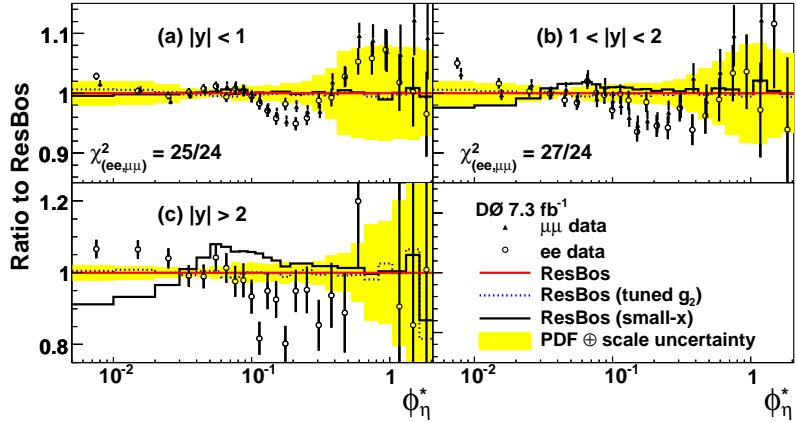}
\label{Fig:phistar}
\caption{
Ratio of measured $(1/\sigma) (d\sigma/d\phistar)$, and alternative ResBos predictions, to the nominal ResBos prediction.
The yellow band around the ResBos prediction represents the uncertainty due to renormalisation and factorisation
scale variation added in quadrature with PDF parameter variations.}
\end{figure}

The D\O\ Collaboration recently measured $(1/\sigma)$ $(d\sigma/d\phistar)$
using 7.3 fb$^{-1}$ of data, in the $e^+e^-$ and $\mu^+\mu^-$ decay channels,
and in three bins of dilepton rapidity~\cite{Dzero_phistar}.
The measured distributions are compared to predictions from the ResBos program
in figure~\ref{Fig:phistar}, with a modest level of agreement.

ResBos includes a non-perturbative form factor which has been tuned to
simultaneously describe low-$Q^2$ Drell-Yan data, and Tevatron Run I $Z/\gamma^*$
data~\cite{BLNY}.
Floating the $g_2$ parameter, which controls the width of the form factor, 
does not substantially improve the agreement, as represented by the blue line
in figure~\ref{Fig:phistar}.
Recently, the $x$-dependence of the non perturbative form factor has received
some attention, and an additional ``small-$x$ broadening'' was suggested to
describe SIDIS data from HERA~\cite{smallx},
which would have significant effects at the LHC~\cite{smallx_LHC}.
The $|y| > 2$ data clearly disfavour the small-$x$ modification, 
which is represented by the black line in figure~\ref{Fig:phistar}.

\section{\boldmath{$Z/\gamma^*$} lepton angular distributions and forward-backward asymmetry}

The angular distributions of the leptons from $Z/\gamma^*$ decays
are often considered in the Collins-Soper frame~\cite{cs_frame},
and are predicted by perturbative QCD~\cite{angles_theory} to take the following form:

\begin{centering}
\medskip
\begin{tabular}{ r c l }

\( \frac{d\sigma}{d\cos\theta d\phi} \) & \( \propto \) & \( (1+\cos^2\theta) \) \\
 & + & \(  \frac{1}{2}A_0(1-3\cos^2\theta) + A_1\sin 2\theta \cos\phi \) \\
 & + & \(  \frac{1}{2}A_2\sin^2 \theta \cos 2\phi + A_3 \sin\theta\cos\phi \) \\
 & + & \(  A_4\cos\theta + A_5\sin^2\theta \sin 2\phi \) \\
 & + & \(  A_6\sin 2\theta \sin \phi + A_7 \sin \theta \sin \phi \) \\
 \end{tabular}
\medskip
\end{centering}

where $\theta$ and $\phi$ are the polar and azimuthal angles respectively~\cite{cs_frame}.
The coefficients, $A_i$, are dependent on the kinematics of the dilepton system;
in particular the $p_T$.
The $A_5, A_6, A_7$ parameters are calculated to be negligible~\cite{angles_theory}.
The $A_4(\cos\theta)$ term generates an asymmetry in the $\cos\theta$ distribution,
and is due to the different couplings of the $Z$ boson to left- and right-handed
fermions, whose relative strength is determined by the value
of $\sin^2\theta_W$.


The forward-backward asymmetry is defined as
$A_{FB} = (\sigma_F - \sigma_B)/(\sigma_F + \sigma_B)$ 
, where $\sigma_F$ and $\sigma_B$ are the cross sections for forward ($\theta > 0$) and backward ($\theta < 0$) events
respectively. 
Interference between the $Z$ and the $\gamma^*$ diagrams leads to an enhanced asymmetry
for masses away from the $Z$ pole.
At higher invariant masses, $A_{FB}$ is sensitive to the presence of additional gauge bosons.
$A_{FB}$ is sensitive to the couplings of the light quarks to the $Z$, which are relatively
poorly constrained by measurements at LEP.

\begin{figure}
\centering
\includegraphics[width=0.4\textwidth]{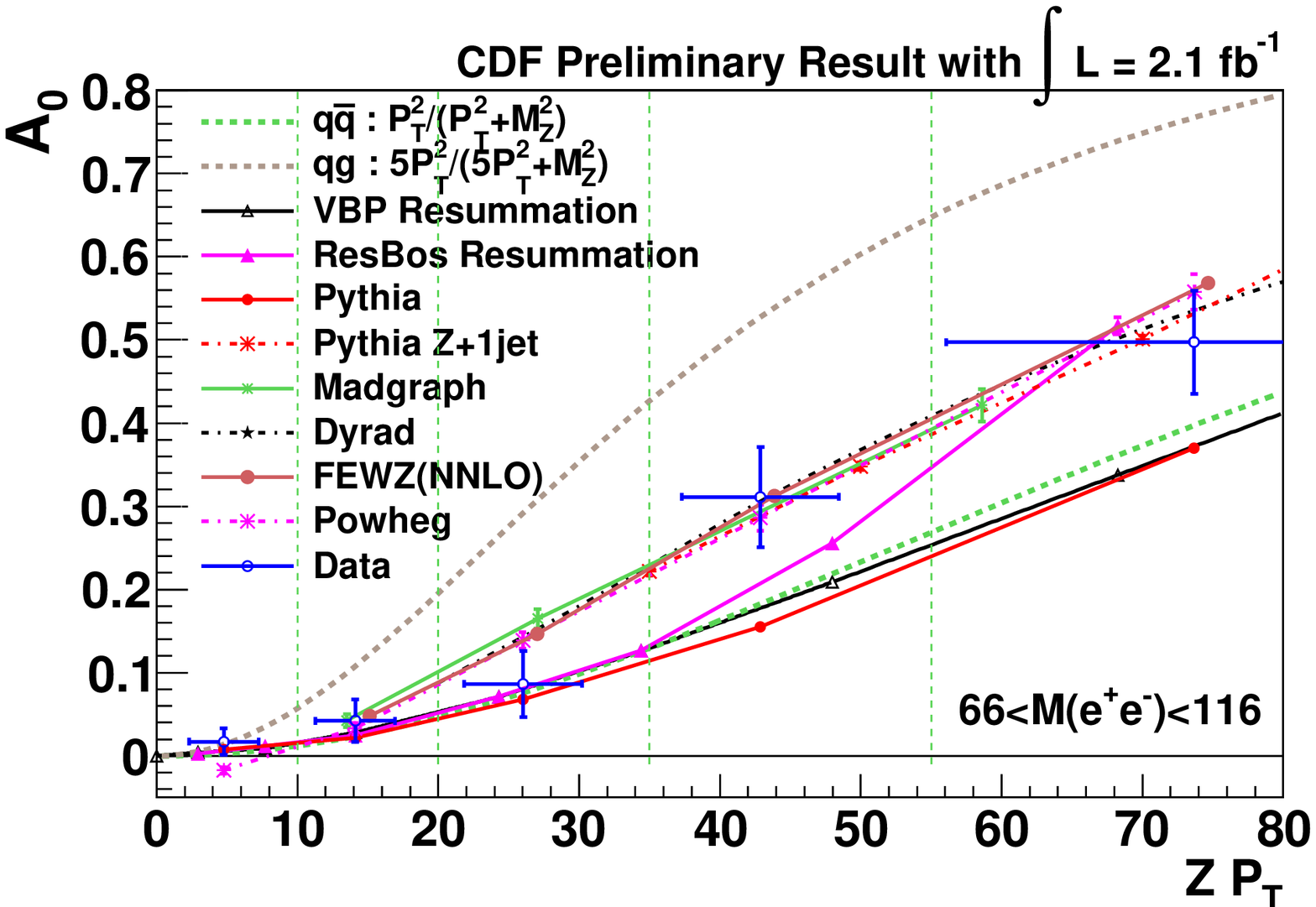}
\includegraphics[width=0.4\textwidth]{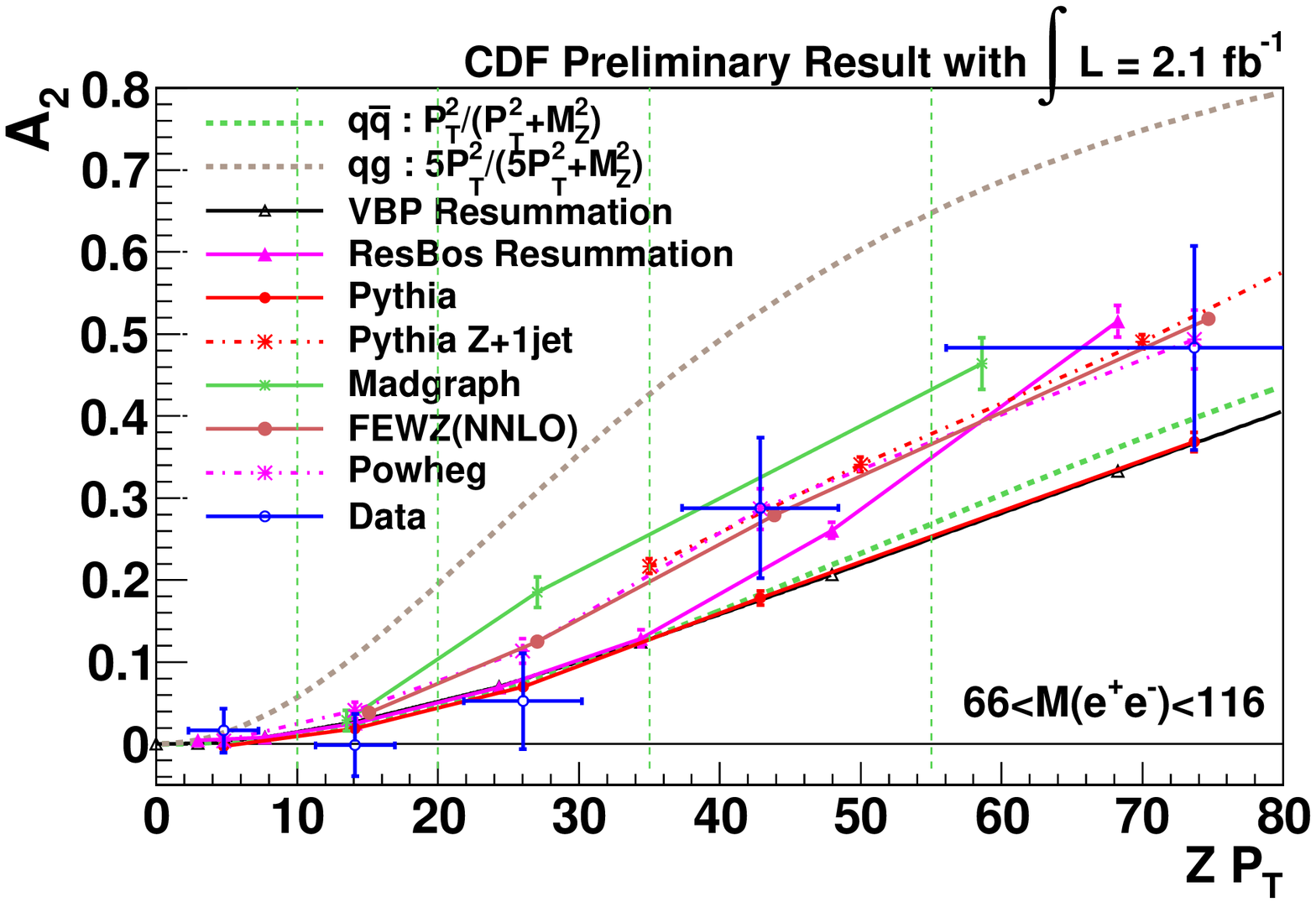}
\includegraphics[width=0.4\textwidth]{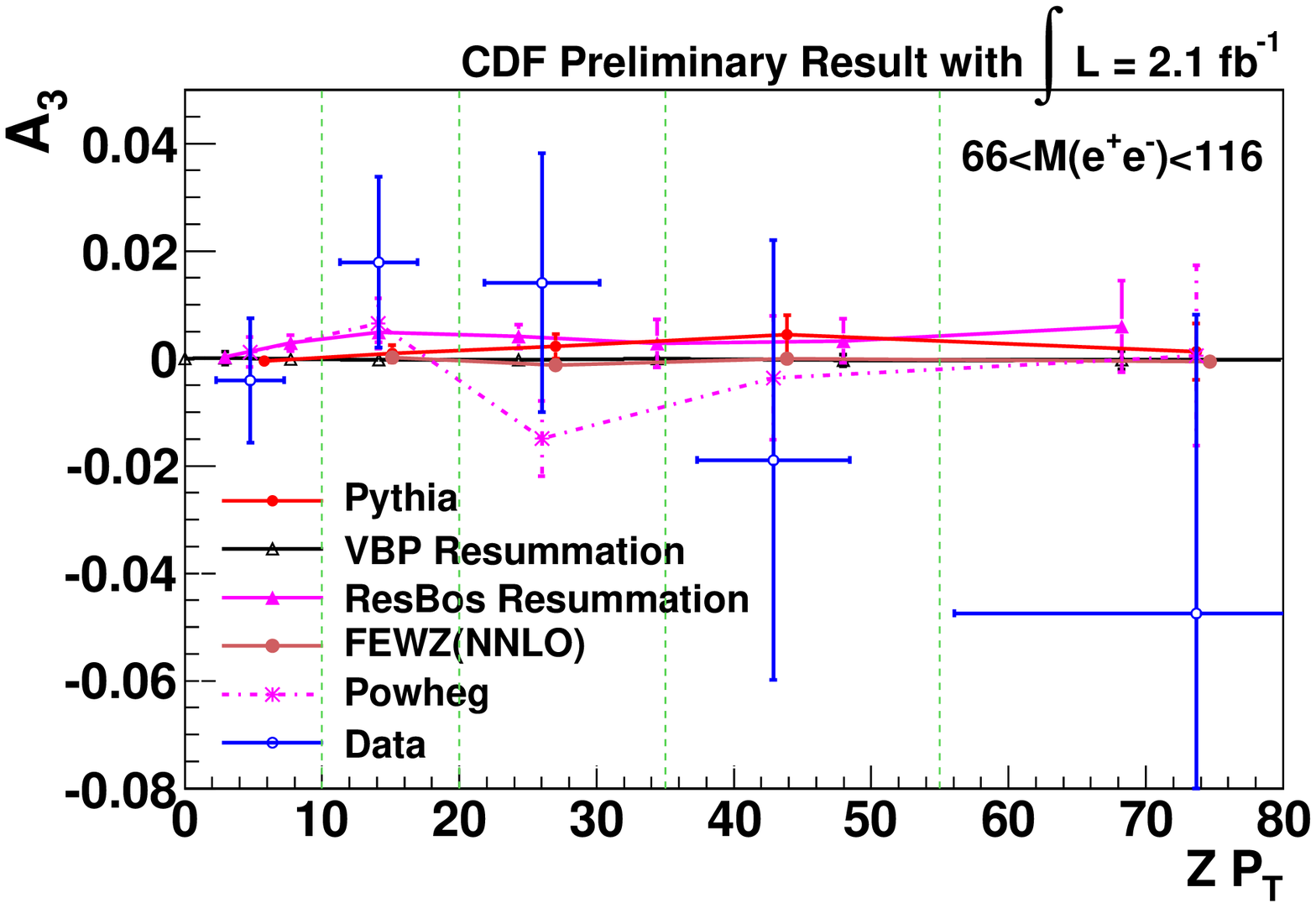}
\includegraphics[width=0.4\textwidth]{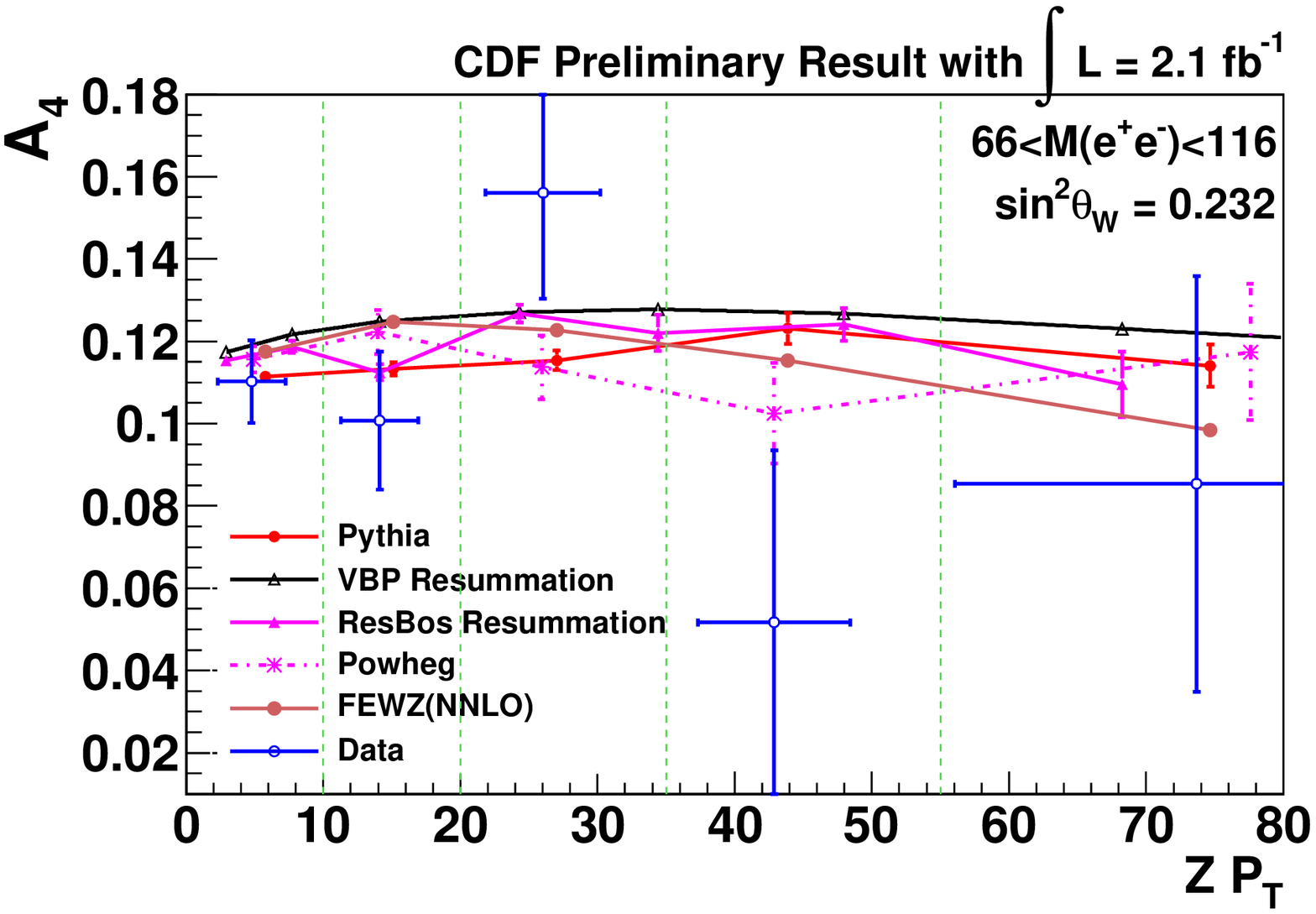}
\label{Fig:Angles}
\caption{Measured angular coefficients as a function of dilepton $p_T$, compared
to various QCD predictions.}
\end{figure}

The CDF collaboration have measured $A_0$, $A_2$, $A_3$ and $A_4$ 
as a function of the dilepton $p_T$, using 2.1 fb$^{-1}$ of data in the $e^+e^-$ decay
channel~\cite{CDF_angles}.
The data are compared to various QCD predictions in figure~\ref{Fig:Angles}.
The $A_4$ parameter (multiplying the $\cos\theta$ term) is directly related to the
$A_{FB}$, and thus also the value of $\sin^2\theta_W$.
The $A_4$ measurement is translated into a measurement of $\sin^2\theta_W = 0.2329 \pm 0.0008 ^{+0.001}_{-0.0009}$,
where the first uncertainty is experimental and the second is theoretical.


\begin{figure}
\centering
\includegraphics[width=0.4\textwidth]{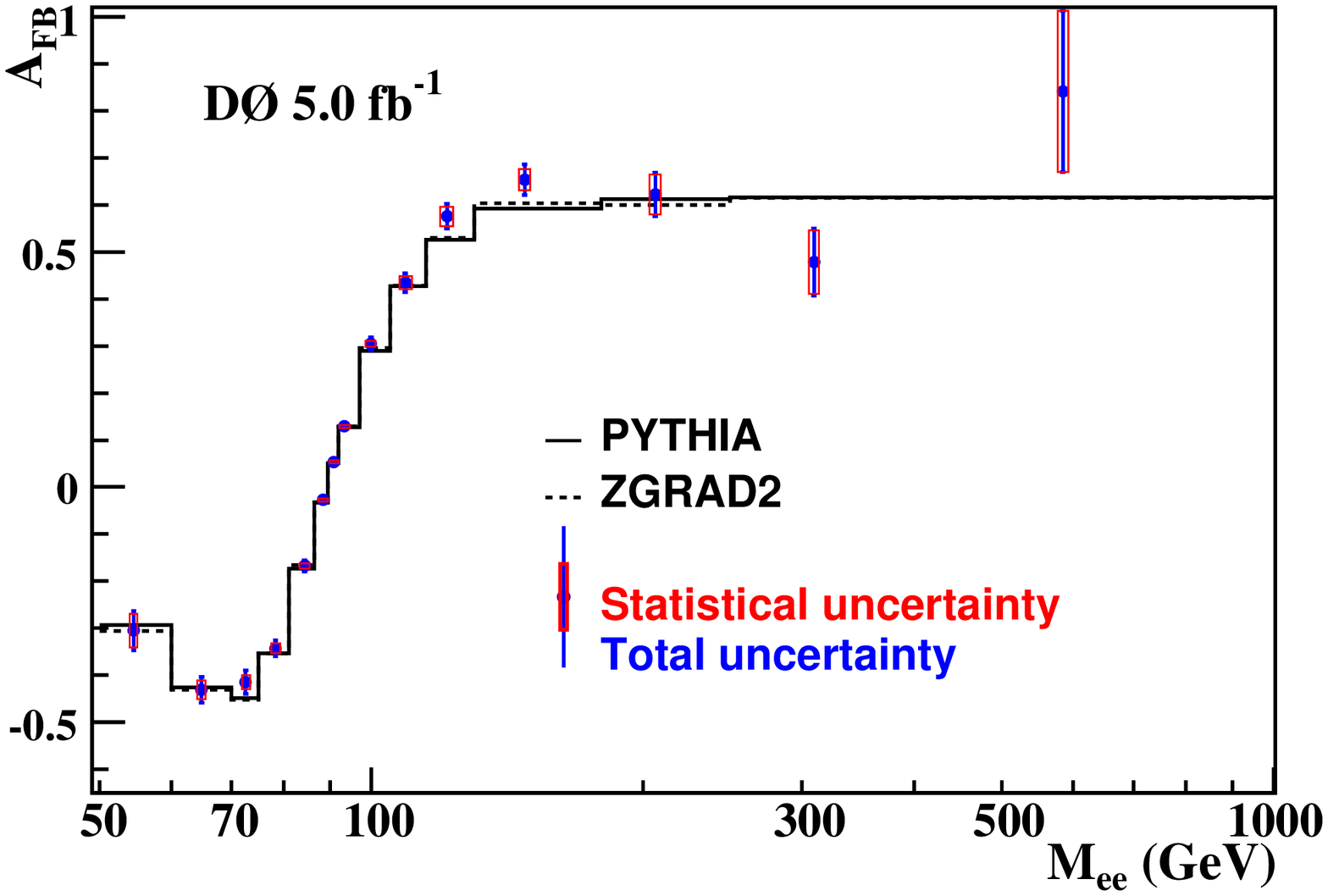}
\includegraphics[width=0.30\textwidth]{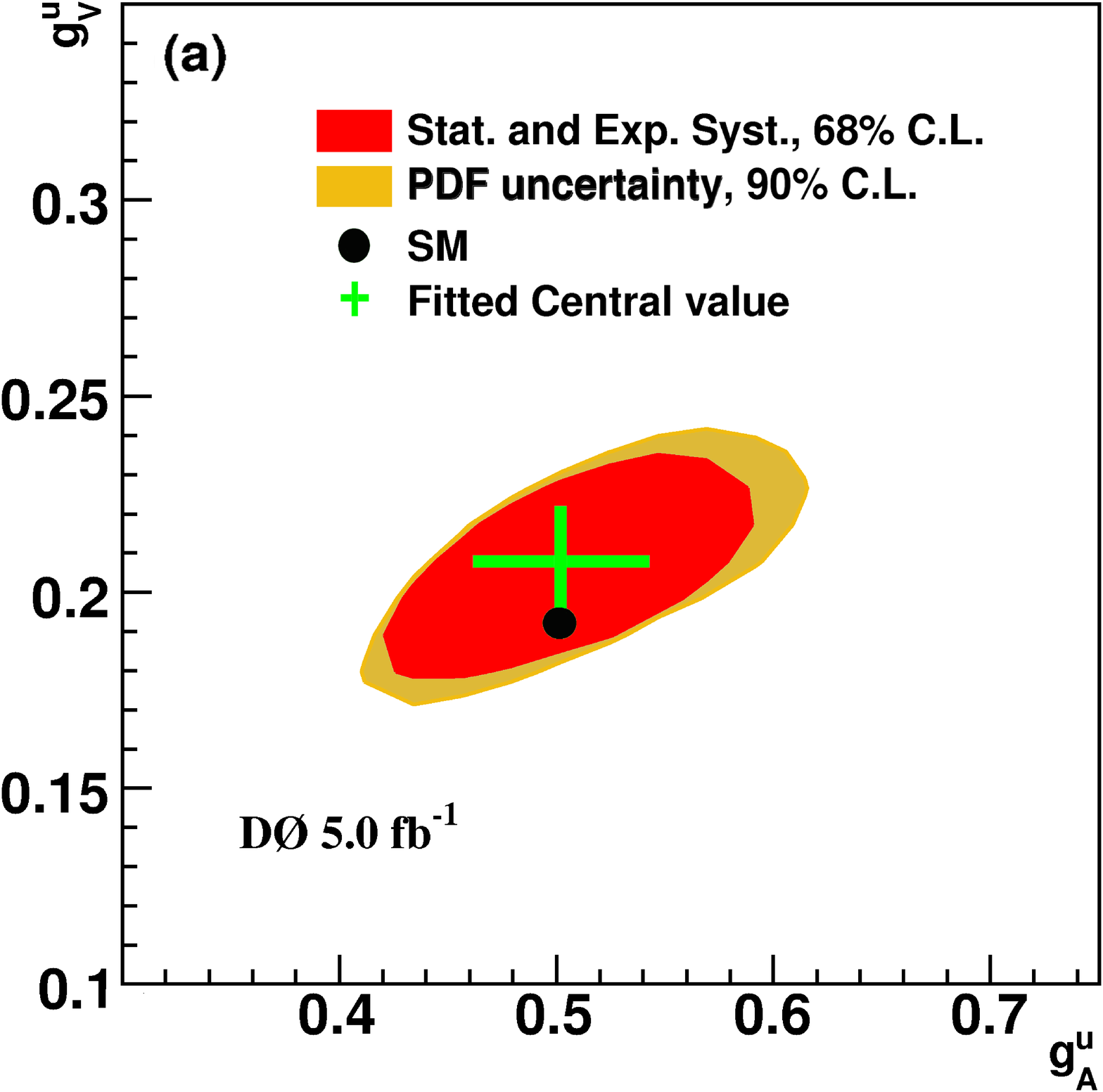}
\includegraphics[width=0.28\textwidth]{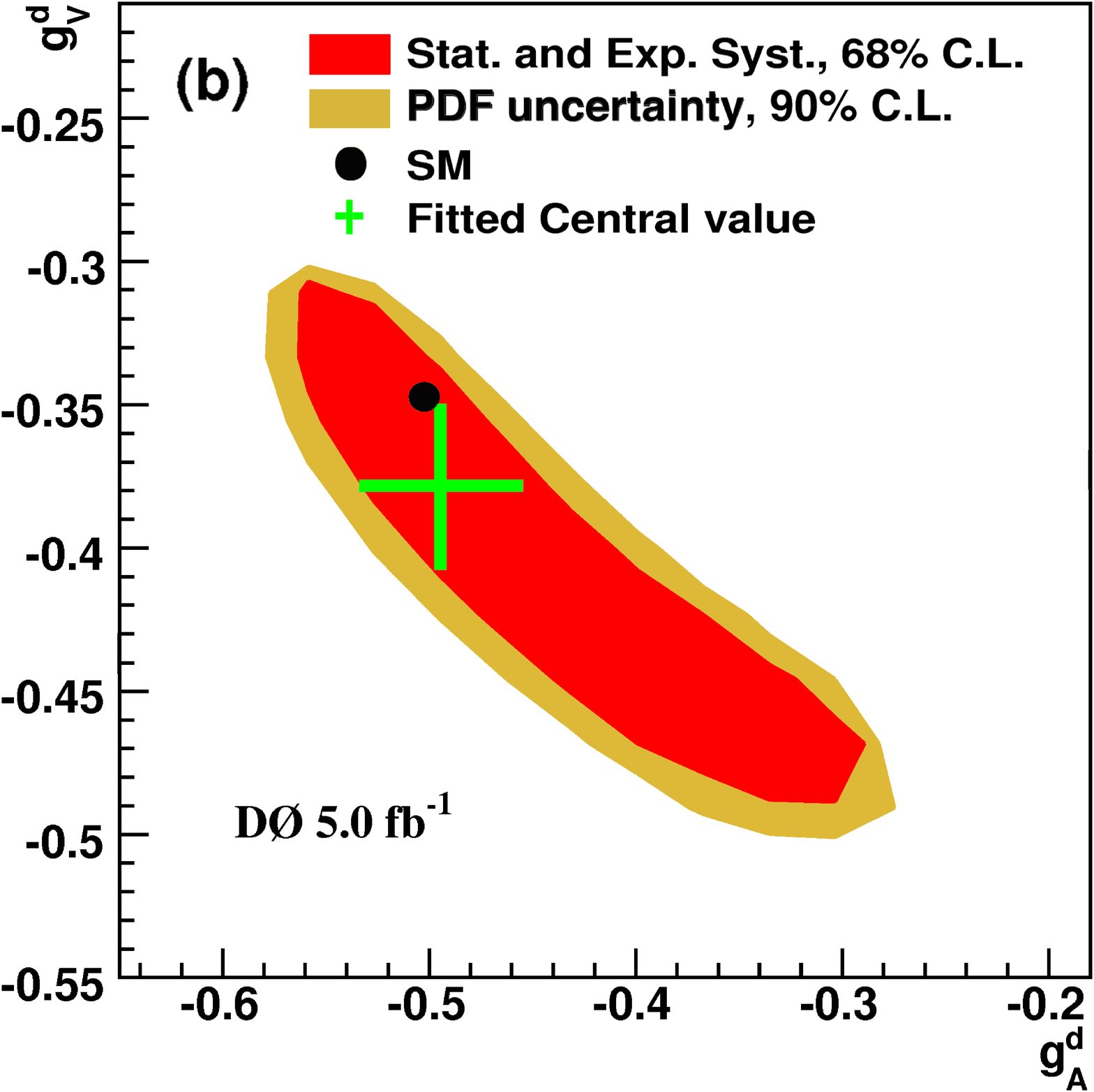}
\label{Fig:AFB_Dzero}
\caption{Left: measured $A_{FB}$ compared to Standard Model predictions. Middle and right: measured $u$ and $d$ quark couplings to the $Z$.}
\end{figure}

The D\O\ Collaboration has recently measured $A_{FB}$ as a function of the dilepton invariant
mass, using 6.1 fb$^{-1}$ of data, in the $e^+e^-$ channel~\cite{Dzero_AFB}.
Figure 6 shows that the measurement is in reasonable agreement 
with Standard Model predictions.
In addition, the couplings of the $u$ and $d$ quarks to the $Z$ are extracted
as shown in figure 6.
A value of $\sin^2\theta_W$ is extracted as 0.2309 $\pm$ 0.001,
in good agreement with the world average.

\section{Conclusions}

Recent analyses of single $W$ and $Z$ boson production properties and asymmetries
from the CDF and D\O\ experiments at the Fermilab Tevatron are presented.
A measurement of the $W$ boson production asymmetry in $W$ $\rightarrow$ $e\nu_{e}$ events
from CDF is in good agreement with QCD predictions.
Conversely, a measurement of the muon charge asymmetry in $W$ $\rightarrow$ $\mu\nu_{\mu}$ events
from D\O\ is in modest agreement with QCD predictions.
The $Z/\gamma^*$ production cross section, and rapidity distribution is measured in the
$e^+e^-$ decay channel by CDF, and agrees well with QCD predictions.
The shape of the $Z/\gamma^*$ transverse momentum distribution is measured using
1 fb$^{-1}$ of data in the $\mu^+\mu^-$ decay channel by D\O,
in reasonable agreement with various QCD predictions.
The \phistar\ variable was recently proposed as an alternative
variable for studying the transverse momentum.
A measurement of the shape of the \phistar\ distribution from D\O\ using
7.3 fb$^{-1}$ of data, in the $e^+e^-$ and $\mu^+\mu^-$ decay channels
is in modest agreement with a state-of-the-art QCD prediction.
Four coefficients describing the angular distributions of the decay leptons
from $Z/\gamma^*$ decays are studied
in the $e^+e^-$ channel by CDF using 2.1 fb$^{-1}$ of data.
D\O\ measures $A_{FB}$ as a function of the
dilepton invariant mass, using 5 fb$^-1$ of data in the $e^+e^-$ decay channel,
in agreement with a QCD prediction.
This measurement is used to extract $\sin^2\theta_W = 0.2309 \pm 0.001 $,
and the most precise determination of the $Z$ boson couplings to $u$ and $d$ quarks.

\end{document}